\documentclass[twocolumn,amsmath,amssymb,showpacs,pre]{revtex4-1}
\usepackage[export]{adjustbox}
\usepackage{subcaption}
\usepackage{mwe}
\usepackage{ragged2e}
\usepackage{overpic}

\usepackage{amsmath}
\usepackage{helvet}
\usepackage{courier}
\usepackage{amssymb}
\usepackage{amsthm}
\usepackage[T1]{fontenc}
\usepackage[latin9]{inputenc}
\usepackage{makeidx}         
\usepackage{graphicx}        
\usepackage[bottom]{footmisc}

\usepackage{hyperref}
\hypersetup{colorlinks=true,
  linkcolor=black,
  anchorcolor=black,
  citecolor=red,
  urlcolor=brown,
  pdfauthor={Adriano Barra and Antonio Moro}}

\providecommand{\be}{\begin{equation}}
  \providecommand{\ee}{\end{equation}}
\providecommand{\bea}{\begin{eqnarray}}
  \providecommand{\eea}{\end{eqnarray}}
\providecommand{\beas}{\begin{eqnarray*}}
  \providecommand{\eeas}{\end{eqnarray*}}

\providecommand{\beni}{\begin{equation*}}
  \providecommand{\eeni}{\end{equation*}}

\providecommand{\bw}{\begin{widetext}}
  \providecommand{\ew}{\end{widetext}}

\def\be{\begin{equation}}
\def\ee{\end{equation}}
\def\bfig{\begin{figure}[htb]}
\def\efig{\end{figure}}

\newcommand{\e}[1]{\,{\rm e}^{#1}\,}

\def\Tr{{\operatorname{Tr\,}}}

\newcommand{\benumerate}{\begin{enumerate}}
\newcommand{\eenumerate}{\end{enumerate}}

\newcommand{\der}[2]{\frac{\partial #1}{\partial #2}}

\newcommand{\caH}{{\mathcal H}}

\date{}

\begin{document}

\author{Costanza Benassi  and  Antonio Moro \\
~\\
\footnotesize{Department of Mathematics, Physics and Electrical Engineering, Northumbria University Newcastle\\
Newcastle upon Tyne, United Kingdom }}

\title{Thermodynamic Limit and Dispersive Regularisation in Matrix Models}

\date{\today}

\begin{abstract}
We show that Hermitian matrix models support the occurrence of a new type of phase transition characterised by dispersive regularisation of the order parameter near the critical point. Using the identification of the partition function with a solution of a reduction of the Toda hierarchy, known as Volterra system, 
we argue that the singularity is resolved by the onset of a multi-dimensional dispersive shock \textcolor{black}{of the order parameter} in the space of coupling constants. This analysis explains the origin and mechanism leading to the emergence of {\it chaotic behaviours} observed in $M^{6}$ matrix models and extends its validity to even nonlinearity of arbitrary order.
\end{abstract}

\maketitle

\section{Introduction}
Random matrix models, originally introduced as an attempt to model the complex structure of heavy nuclei energy spectra, have since become a universal paradigm for  modelling complex phenomena. They naturally arise in connection with different areas of mathematics and physics, from quantum field theory to the theory of \textcolor{black}{completely integrable dynamical systems}~\cite{wigner,dysonI, mehta}. 

For instance, in the context of quantum field theory, Hermitian matrix models can be introduced as a discrete approximation of 2D quantum gravity. Based on this result, a celebrated conjecture of Witten~\cite{witten}, proven by Kontsevich~\cite{konts}, establishes the \textcolor{black}{connection} between the free energy of 2D quantum gravity \textcolor{black}{and a particular solution of the 
Korteweg-de Vries (KdV) equation. The KdV equation, the prototypical example of soliton equation \cite{ablowitz}, first appeared in water wave dynamics to model small amplitude elevation waves in shallow water. 
Along with solitons, another important class of solutions of the KdV equation is constituted by 
Dispersive Shock Waves (DSW)~\cite{whitham,gurevich-pitaevskiy}. DSWs occur as a universal regularisation mechanism of singularities in dispersive hydrodynamics, and effectively explain the emergence of a variety of complex behaviours in hydrodynamic systems~\cite{whitham,gurevich-pitaevskiy,el, eh, ehs,congy}.
The remarkable mathematical structure of completely integrable systems, of which the KdV equation is possibly the most celebrated example, such as the existence of an infinite number of conservation laws, allows for an effective and detailed description of dispersive shock waves~(see e.g. \cite{gurevich-pitaevskiy,el, eh, ehs}). The above mentioned Witten's conjecture established an unexpected connection between quantum field theory and dispersive hydrodynamics via the identification of correlators of 2D gravity with conserved densities of a particular solution of the KdV equation}. Thereafter, similar correspondence between different matrix models on Hermitian, Unitary, Symmetric and Symplectic ensembles and \textcolor{black}{completely integrable dynamical systems} have been discovered (see e.g. \cite{gerasimov,adler,adler2,okounkov,takasaki} and references therein). \textcolor{black}{The relevance of these connections lies in the fact that methods developed for solving completely integrable dynamical systems are indeed effective for the study of the associated matrix models and unveil further intriguing mathematical structure and elegance.} Moreover, extensive studies of properties of partition functions for matrix models \textcolor{black}{led to remarkable} connections between the theory of integrable systems, statistical mechanics, quantum field theory, algebraic and enumerative geometry \cite{witten,adler,okounkov,takasaki,dubrovin,dubrovin2}. 
In this paper  \textcolor{black}{we investigate Hermitian matrix models and observe a novel type of  phase transition}, resulting from a dispersive regularisation mechanism of the order parameter in the space of coupling constants and the consequent onset of a dispersive shock. We provide the physical conditions and scaling regime under which this phenomenon occurs. In particular, we find that, unlike classical mean field models, multivalued solutions at the leading order of the asymptotic expansion do not necessarily correspond to a phase transition.
For the sake of simplicity, we focus on a case of Hermitian Matrix Model with even nonlinear interaction terms and  \textcolor{black}{and its formulation in terms of the 1D Toda hierarchy.} However, our considerations can be extended to other matrix ensembles. \textcolor{black}{The Toda Lattice is an important example of completely integrable nonlinear dynamical system which, in the continuum limit, contains as a particular case several examples of soliton equations, including the KdV equation}.  We also note that asymptotic properties of partition functions in the thermodynamic limit of one-matrix models with even and odd nonlinearity and their relation with the Toda \textcolor{black}{Lattice} have been previously considered in \cite{ercolani1,ercolani2}.  
A key point in our analysis is that the sequence of partition functions $Z_{n}$ for the one-matrix model of $n \times n$ matrices  can be expressed in terms of a particular solution of a suitable restriction of the Toda Lattice equations, \textcolor{black}{known as Volterra Lattice (or discrete KdV equation)}. 
The complete integrability of the Volterra Lattice system implies the existence of infinitely many conservation laws. The Volterra Lattice together with its set of symmetries associated to the conservation laws constitutes the Volterra hierarchy. We show that the identification of the Volterra hierarchy with the matrix model is based on a one to one correspondence between coupling constants and equations of the hierarchy. Each equation of the hierarchy provides the evolution of the order parameter of the theory as a function of the associated coupling constant, that is identified with the time variable of the chosen equation. The partition function $Z_{n}$ is therefore specified by the state of the $n-$th point of the lattice \textcolor{black}{for the corresponding values of the coupling constants (i.e. time parameters of the hierarchy)}. Most importantly, the dynamics \textcolor{black}{on the lattice} is uniquely specified by the initial conditions that are given in terms of the partition function of the {\it free} model, i.e. where all coupling constants vanish. 
In this respect, the model is relatively simpler than the case of 2D gravity studied in~\cite{witten} where the initial condition is specified by additional symmetries that are compatible with the hierarchy, namely the Virasoro constraints \cite{dubrovin2,adler}.  

\textcolor{black}{We thus exploit the relation between the matrix model under consideration and the Volterra Lattice in order to explore the phase diagram of the model}.
In his pioneering work \cite{jurkiewicz2},  Jurkiewicz observed that a natural order parameter can be introduced by using orthogonal polynomial decompositions and combinatorial considerations \cite{ParisiEtAl}. Such order parameter develops, in the thermodynamic limit, a singularity that is regularised by oscillations revealing an apparent underlying chaotic behaviour of the system \cite{jurkiewicz2, senechal}. Rigorous proof of the occurrence of asymptotic oscillations of the partition function has been found in~\cite{ken,grava}. 
In this work we explain the oscillatory behaviour of the order parameter as the occurrence of a dispersive shock. Using the fact that the thermodynamic limit ($n\rightarrow \infty$)  of the random matrix model is equivalent to the continuum limit of the Volterra Lattice, given by a system of partial differential equations of hydrodynamic type \cite{deift,kodama, michor}, we show that the order parameter evolves in the space of the coupling constants as a shock wave solution of the associated hydrodynamic system. 
The chaotic phase is therefore interpreted as a dispersive shock propagating through the chain in the continuum/thermodynamic limit. The intriguing complexity of its phase diagram can hence be explained in the context of dispersive hydrodynamics. The physical constraint on the signature of the order parameter determines whether a  catastrophe evolves into a dispersive shock.

Considerations above outline the following \textcolor{black}{general} scenario: when a thermodynamic system undergoes a phase transition, some specific quantities, the order parameters, develop singularities. In the context of conservation laws of hydrodynamic type, when a singularity (hydrodynamic catastrophe) occurs, viscosity and dispersion underpin two different mechanisms of regularisation of such singularity. In presence of small viscosity the solution develops a sharp but smooth wavefront \cite{whitham}. If small viscosity is replaced by small dispersion, when the wavefront approaches the point of gradient catastrophe the dispersion induces initially small oscillations that further evolve into a dispersive shock~\cite{gurevich-pitaevskiy, el,ehs}. In classical mean field fluid and spin models phase transitions are associated to classical shocks of order parameters in the space of thermodynamic parameters \cite{moro1, barra, genovese}. In this work we show that the chaotic behaviour observed in~\cite{jurkiewicz2} is indeed a phase transition where the order parameter develops a singularity that is resolved via dispersion rather than viscosity as in classical spin models. \textcolor{black}{The observation that phase transitions in matrix models are explicitly connected to the occurrence of a dispersive shock in the order parameter paves the way to the application of current methods of dispersive hydrodynamics in the context of quantum field theories.}\\

\section{Hermitian Matrix Model}
We study the model defined by the partition function
\be
\label{partition}
Z_n(\mathbf{t}) = \int_{\caH_n} \e{H(M)} dM,
\ee
where $M$ are Hermitian matrices of order $n$,
\textcolor{black}{
\[H(M) =\Tr \left (-M^{2}/2 + \sum_{j=1}^\infty t_{2j} M^{2j} \right)\]}is the Hamiltonian, with $\mathbf{t} = \{t_{2j}\}_{j\geq 1}$ the coupling constants,  and $dM$ is the Haar measure in the space of Hermitian Matrices ${\caH_n}$.  Based on a classical result by Weyl~\cite{weyl}, the partition function~(\ref{partition}) is proportional to an integral over the eigenvalues of the matrix $M$, that is $Z_{n}(\mathbf{t}) = c_{n} \tau_n(\mathbf{t})$ where $c_{n}$ is a constant and
\be
\tau_n(\mathbf{t}) = \frac{1}{n!}\int_{\mathbb{R}^n}\Delta_{n}(\boldsymbol{\lambda})^{2} \prod_{i=1}^n \left( \e{H(\lambda_{i})}d\lambda_i \right)
\label{eqtau}
\ee
where  $\Delta_n(\boldsymbol{\lambda}) = \prod_{1\leq i < j \leq n } (\lambda_{i} - \lambda_{j})$ is the Vandermonde determinant. 
A theorem by Adler and van Moerbeke~\cite{adler} implies that the quantity~\eqref{eqtau} can be interpreted as a tau-function of the Toda hierarchy restricted to the even flows
\be
\der{L}{t_{2k}} = \left [ \frac{1}{2}\left(L^{2k} \right )_{s}, L \right ] \quad k =1,2,\dots .
\label{eqTodaHierarchy}
\ee
with $L$ the tridiagonal symmetric  Lax matrix of the form
\begin{equation}
L = \left(
\begin{matrix}
0 & b_{1} & 0  & 0 & \dots \\
b_{1} & 0 & b_{2}  & 0 & \dots \\
0 & b_{2} & 0  & b_{3} & \dots \\
\vdots & \ddots & \ddots & \ddots & \ddots
\end{matrix}
\right)
\label{laxmat}
\end{equation}
where $b_{i} =   \sqrt{\tau_{i+1}\tau_{i-1}/\tau_i^2}$ and $\left( L^{2 k} \right )_{s}$ denotes the skew-symmetric part of the matrix $L^{2k}$ (see e.g. \cite{adler}). The solution of interest is specified by the initial conditions $ b_i(\mathbf{0}) = \sqrt{n}$ obtained via a direct calculation of Gaussian integrals for the quantities
$\tau_n(\mathbf{0}) = (2\pi)^{n/2}\prod_{j=1}^n  j!/n!\;$. 
We note that the Lax matrix of the type~\eqref{laxmat}, originally considered in~\cite{kac}, and more recently in~\cite{ercolani1}, corresponds to a reduction of the even Toda hierarchy known as Volterra hierarchy. \textcolor{black}{The description of the Volterra hierarchy in terms of the matrix resolvent, its $\tau$ structure and application to ribbon graphs and Hodge integrals has been recently studied in~\cite{Yang}}. Incidentally, we also mention that the model with odd nonlinearities is different from the present case and its relation with the Toda hierarchy has been considered in~\cite{ercolani2}. 

\textcolor{black}{Writing the equations of the hierarchy \eqref{eqTodaHierarchy} in terms of lattice variables $b_{n}$ we have
\be
\label{bneq}
\der{b_n}{t_{2k}} = \frac{b_n}{2}\biggl(b_{n+1}(L^{2k-1})_{n+1, n+2}- b_{n-1} (L^{2k-1})_{n-1,n}\biggr).
\ee
Introducing the notation $B_n = b_n^2$ and $V^{(2k)}_n = b_n (L^{2k-1})_{n, n+1}$ and multiplying both sides by $b_{n}$, the above equation~(\ref{bneq}) reads as follows
\be
\label{Todareduced}
\der{B_n}{t_{2k}} =  B_n (V^{(2k)}_{n+1} - V^{(2k)}_{n-1}) \quad k=1,2,\dots
\ee
One can simply prove by  induction that $V^{(2k)}_n$ is a linear combination of products of the variable $B_n$ evaluated at different points on the lattice.
For instance, for the first three non-trivial flows we have
\begin{align*}
V^{(2)}_n &= \,B_n \\
V^{(4)}_n &= V^{(2)}_n \left (V^{(2)}_{n-1}+V^{(2)}_n+V^{(2)}_{n+1} \right)\\
V^{(6)}_n &= \,V^{(2)}_n \left (V^{(2)}_{n-1}V^{(2)}_{n+1} + V^{(4)}_{n-1} + V^{(4)}_n + V^{(4)}_{n+1} \right).  
\end{align*}
In the following we refer to $B_{n}$ as the {\it order parameter} of the theory.  
}
Based on a result in~\cite{bonora}, we have proven that the required solution to the above reduction of the even Toda hierarchy is given by the recursive formula ({\it string equation})
\be
n = B_n - \sum_{j=1}^\infty 2j\, t_{2j}\, V^{(2j)}_n .
\label{eqrecurrence}
\ee
\textcolor{black}{Indeed, Eq. \eqref{eqrecurrence} follows from the the string equation for the Toda lattice 
\be
\label{string}
\left[L,P \right] = 1
\ee 
where 
\[
P = -\frac{1}{2}L_{s}  +\sum_{k\geq 1} k t_{2k} \left(L \right)^{2k-1}_{s} \; 
\]
by considering its restriction to even times only~\cite{bonora}.}
Eq.~(\ref{eqrecurrence}) allows to evaluate the order parameter of the $M^{2q}$ model for arbitrary $q$ and generalises the formula obtained by Jurkiewicz for $q=3$ \cite{jurkiewicz,jurkiewicz2}. 

\section{Thermodynamic limit}
We analyse the Matrix Model in the large $n$ (thermodynamic) regime via the continuum limit of the solution~\eqref{eqrecurrence} of the reduced Toda hierarchy. Introducing the scale given by a suitable large integer $N$   and the rescaled variables $u_{n} = B_{n}/N$,  $T_{2k} = N^{k-1} t_{2k}$, Eq.~\eqref{eqrecurrence} reads as follows
\be
\label{eqrecurrenceU}
\frac{n}{N} = u_{n} - \sum_{j=1}^{\infty} 2 j T_{2j} W_{n}^{2j}
\ee
where $W_{n}^{2j} = V_{n}^{(2j)}/N^j$. We then define the interpolating function $u(x)$ such that $u_{n} =  u(x)$ for $x =n/N$ and $u_{n\pm1} = u({x\pm \epsilon})$ with the notation $\epsilon = 1/N$.  \textcolor{black}{Expanding in Taylor series for small $\epsilon$, we obtain an ODE as a formal series
\be
\Omega_{\epsilon} = 0
\label{stringformal}
\ee
where $\Omega_\epsilon$ has the following form
\be
\begin{split}
\Omega_\epsilon :=& - x + (1 - 2 T_{2}) u -  12 T_{4} u^{2} - 60 T_{6}  u^{3} \\&+ \epsilon^2\left(
 p_1 u_{xx} +
p_2 u_x^2 \right) \\
 &+  \epsilon^4 \left(q_1 u_{xxxx} + q_2 u_x u_{xxx} + q_3 u_{xx}^2\right) + O(\epsilon^6),\\
\end{split}
\ee
Here for simplicity we have fixed $T_{2k} = 0$ for $k>3$, with $p_i$ and $q_i$ as follows.
\begin{align*}
&p_{1} = - 4 T_4 u - 60 T_6 u^2 &p_{2} = - 30 T_6 u \\
&q_{1} = - \frac{T_{4}}{3}u - 11 T_{6} u^2  &q_{2} = - 22 T_6  u \\
&q_{3} =  -\frac{33 T_{6}}{2}u. & 
\end{align*}
}At the leading order we have the polynomial equation in $u$ of the form
\be
\label{interpu}
\Omega_{0} := - x + (1 - 2 T_{2}) u -  12 T_{4} u^{2} - 60 T_{6}  u^{3}  = 0.
\ee
\textcolor{black}{In order to obtain further insights on the behaviour of the solution of the recurrence equation~(\ref{eqrecurrenceU}) in the thermodynamics limit, it is interesting to study the continuum limit of the Volterra hierarchy~(\ref{Todareduced}). Proceeding as above, i.e. introducing the interpolating function $u(x)$ in~(\ref{Todareduced}) and expanding $u(x\pm \epsilon)$ in Taylor series, one obtains a compatible hierarchy of dispersive partial differential equations of the form
\be
u_{T_{2k}} = \sum_{n=0}^{\infty} \epsilon^n g^{(k)}_{n} \left(u; u_{x},\dots,\partial_{x}^{n}u \right)
\label{Volterracont}
\ee
where functions $g^{(k)}_{n}$ are differential polynomials of $u$.
For instance, the first member of the hierarchy (for $k=1$), which gives the flow with respect to $T_{2}$, takes the following compact form
\be
\label{T2eqsinh}
u_{T_{2}} = 2 u \left [ \frac{1}{\epsilon}  \sinh \left (\epsilon \partial_{x} \right ) \right ] u
\ee
where the operator stays for the formal McLaurin expansion of $\sinh$.
}\textcolor{black}{In the thermodynamic limit, i.e. $\epsilon  \to 0$, equation ~(\ref{T2eqsinh}) gives, at the leading order, the Hopf equation
\be 
u_{T_{2}} = 2 u u_x.
\label{eqhopf}
\ee
Similarly, higher flows in $T_{2k}$ lead to higher members of the so-called Hopf hierarchy 
\be
\label{hopfh}
u_{T_{2k}} = C_{k} u^{k} u_x
\ee
for suitable constants $C_{k}$, the solution of which, for the assigned initial condition, is implicitly given by the polynomial equation~(\ref{interpu}). The effect of the corrections in the parameter $\epsilon$ will be discussed in the following section.}

\begin{figure}[thb]
\begin{center}
    \includegraphics[scale=0.35, frame]{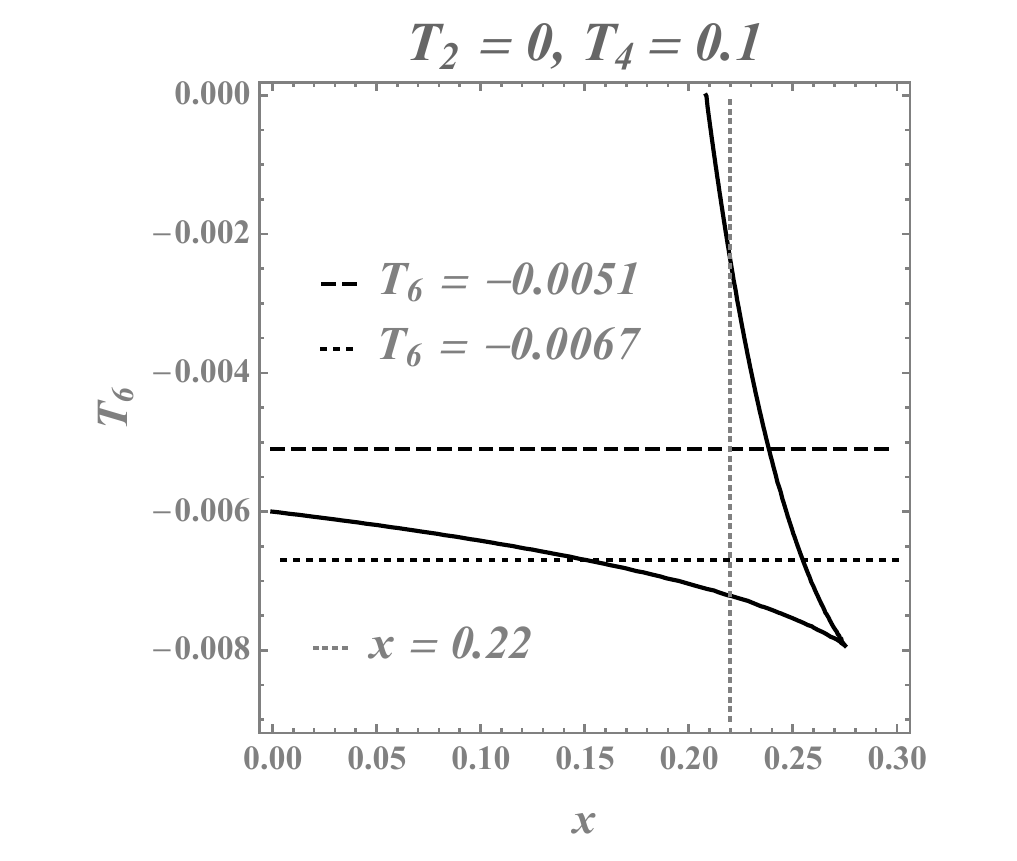}  \caption{\small \textcolor{black}{Critical set $\Delta =0$ in the $x$-$T_6$ plane for $T_2 = 0$ and $T_4 = 0.1$. The convex sector identifies the region $\Delta > 0$ where the equation~\eqref{interpu}  admits multiple roots for the chosen values of parameters. }}
    \label{fig:mixA}
  \end{center}
  \end{figure}

\begin{figure}[thb]
\begin{center}
    \includegraphics[scale=0.35, frame]{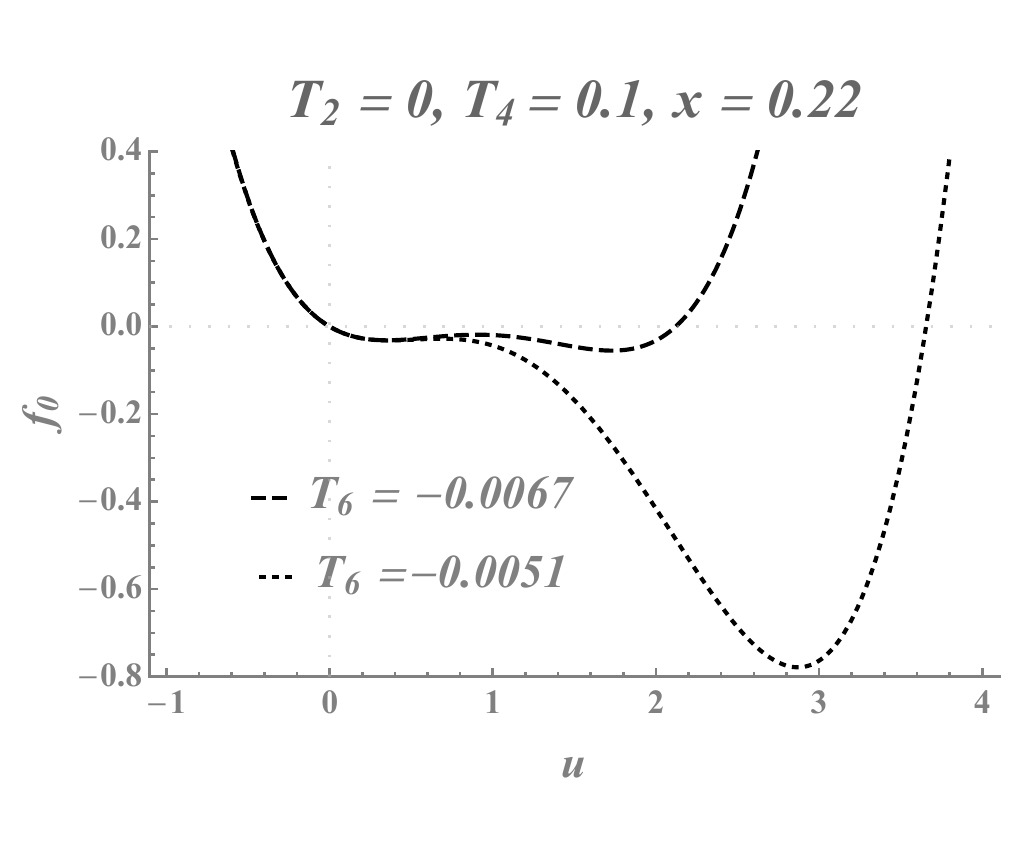}\caption{\small \textcolor{black}{Free energy density for chosen fixed values of the parameters in the region $\Delta > 0$. The chosen values of $T_{6}$ and $x$ correspond respectively to the intersection of the line $x = 0.22$ and the lines $T_{6} = -0.0067$ (thick dashed) and $T_{6} = -0.0051$ (light dashed) as shown in Fig.~\ref{fig:mixA}. } }
    \label{fig:mixB}
  \end{center}
  \end{figure}

\textcolor{black}{For illustrative purposes, we consider the case $t_{2k} = T_{2k} = 0$ for $k>3$, i.e. Eq. \eqref{interpu} is a cubic equation for the order parameter $u$.}
\textcolor{black}{In this case, Eq. \eqref{interpu} provides the condition for extremising the free energy functional $F = \int_{0}^{\beta} f_{0}(u) \; dx$, for some $\beta > 0$, of density
\be
f_{0}[u] = - x u+ \frac{1}{2}  \left(1 - 2 T_2 \right)u^2  - 4 T_4 u^3 - 15 T_6 u^4.
\label{eqFE}
\ee
}
In particular, local minima and maxima depend on the signature of the discriminant $\Delta(x,T_{2},T_{4},T_{6})$ of the cubic equation \eqref{interpu}. If $\Delta >0$ the free energy has two local minima and one local maximum, if $\Delta <0$ the free energy presents one local minimum only. The set in the space of parameters such that $\Delta = 0$ corresponds to the critical set where a phase transition occurs. 
For example, in Fig.~\ref{fig:mixA}  we plot the set $\Delta = 0$ in the $x$-$T_6$ plane for a given choice of $T_2$ and $T_4$ \textcolor{black}{(notice that in order to ensure convergence of the integral \eqref{partition}, we have $T_{6} < 0$)}. The convex sector corresponds to the region where the equation of state \eqref{interpu} admits three real solutions that correspond to the stationary points of the free energy density. \textcolor{black}{Fig.~\ref{fig:mixB} shows the free energy density as a function of $u$ for two different values of $T_{6}$, with fixed $T_2$, $T_4$ and  $x$. For these values of the parameters the discriminant of the cubic is positive, and the free energy has three extremal points. We see that} for $T_{6} = -0.0051$ the difference of the values of the free energy density at its local minima is particularly pronounced compared with the case $T_{6} = -0.0067$. 

\textcolor{black}{The above scenario is compatible with the well known fact that the generic solution of the Hopf hierarchy~(\ref{hopfh}) develops singular behaviour for finite value of the ``time'' variables $T_{2k}$~\cite{whitham}. In the next section, we study these singularities in relation to the occurrence of phase transitions. A phase transition is associated to the occurrence of a dispersive shock induced by dispersive corrections to the Hopf hierachy. Eq.~\eqref{stringformal} obtained as the continuum limit of \eqref{eqrecurrenceU} provides quasi-trivial deformations of the Hopf hierarchy and the behaviour near the singularity that are universally described by a solution of the fourth order analogue of the Painlev\'e I equation \cite{dubrovin3a,dubrovin3,dgkm}.}
\\
\section{Dispersive regularisation}
\textcolor{black}{In the thermodynamic limit, the evolution of the order parameter in the space of coupling constants is governed by Eq. \eqref{hopfh} provided derivatives of $u$ are bounded. In the vicinity of the gradient catastrophe, dispersive corrections in the equation~(\ref{Volterracont}) induce fast oscillations responsible for a rich and interesting phenomenon known as dispersive shock wave~\cite{gurevich-pitaevskiy,eh}}.
\begin{figure}[hbt]
\centering
\begin{subfigure}[b]{0.35\textwidth}
    \includegraphics[width=\textwidth, frame]{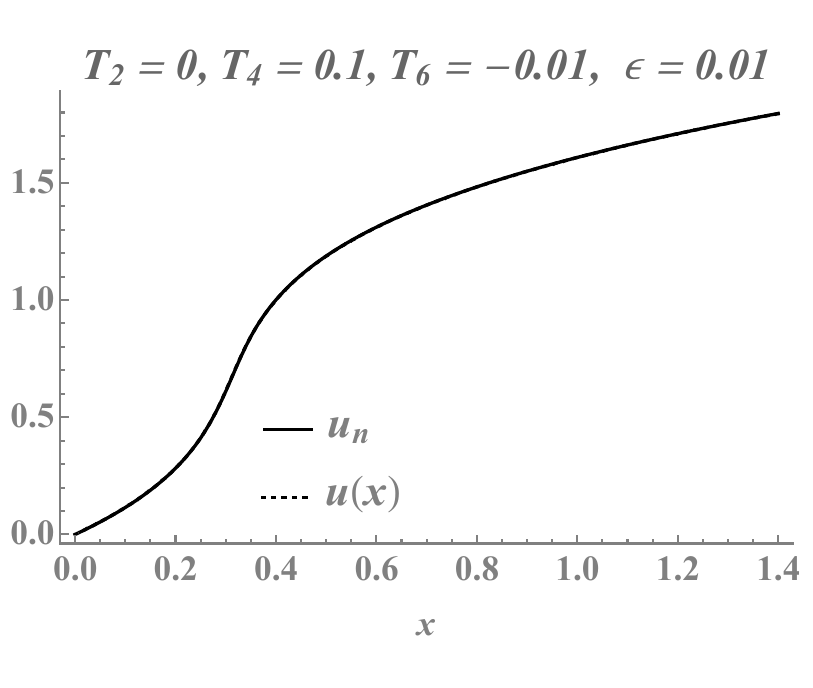}   
  \caption{}
    \label{fig:D}
  \end{subfigure}
 \quad
  \begin{subfigure}[b]{0.35\textwidth}
    \begin{overpic}[width=\textwidth, frame]{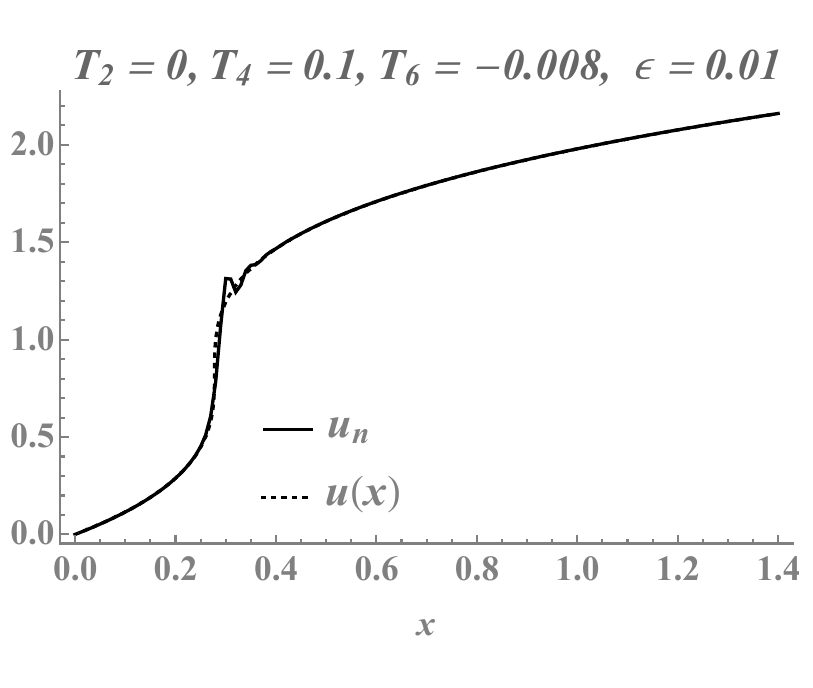}
  \put(52,22){
\frame{\includegraphics[scale=0.16]{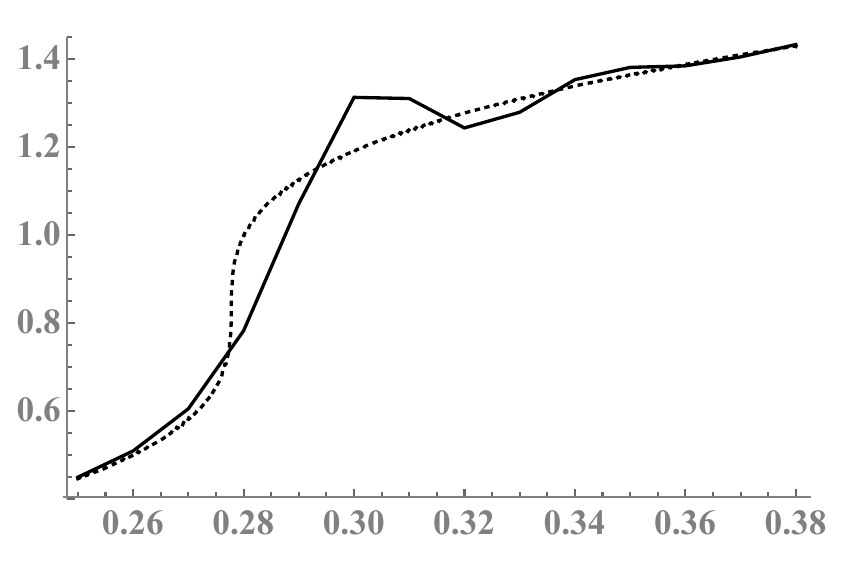}}}
\end{overpic}
  \caption{}
    \label{fig:C}
  \end{subfigure}\\
\caption{\small Comparison of the order parameter evaluated using Eq.~\eqref{eqrecurrenceU} and Eq.~\eqref{interpu} at $T_2 = 0$, $T_4 = 0.1$. In Fig. \ref{fig:D} $T_6 = -0.01$ and $\Delta<0$ for all $x$. In Fig. \ref{fig:C} $T_{6} = -0.008$ and $\Delta = 0$ at \textcolor{black}{the point of gradient catastrophe $x = 5/18 \simeq 0.28$ of the solution $u$ of the equation~\eqref{interpu}.}} The inset  zooms-in around $x=5/18$. \textcolor{black}{As $T_{6}$ increases we observe a steepening of the profile of the order parameter $u_{n}$ and the onset of oscillations in the vicinity of the point of gradient catastrophe. }
  \label{fig:comparison}
\end{figure}
\textcolor{black}{In this section we} illustrate the general phenomenology by considering the particular case $T_{2k}$~$=$~$t_{2k}$~$= 0$ for all $k>3$ so that $T_{2}$, $T_{4}$ and $T_{6}$ are the only non zero coupling constants. This choice allows for a simple but sufficiently general analysis demonstrating as {\it chaotic behaviours} observed in~\cite{jurkiewicz} correspond to a type of phase transition comprised by a dispersive shock of the order parameter. The shock occurs as a dispersive regularisation mechanism of a particular solution of the hierarchy~\eqref{Todareduced} in the continuum limit. 

In Fig.~\ref{fig:comparison} we compare the order parameter $u(x)$ obtained as solution of the recurrence equation~\eqref{eqrecurrenceU} and the limit equation~\eqref{interpu}. Values $T_{2}$, $T_{4}$ and $T_{6}$ are chosen in such a way that the solution of the cubic equation~\eqref{interpu} is single valued. Fig.~\ref{fig:D} shows that the two solutions fully overlap for sufficiently small $\epsilon$, but, as shown in Fig.~\ref{fig:C}, a relevant deviation is observed in the vicinity of the point of gradient catastrophe of the solution to Eq.~\eqref{interpu}.


Figs.~\ref{fig:mixC} and~\ref{fig:mixD} show a comparison between the cubic solution \eqref{interpu} and the exact solution~\eqref{eqrecurrenceU} for different values of $T_6$ within the convex region shown in Fig.~\ref{fig:mixA}, where the solution of~\eqref{interpu} is multivalued. Both figures demonstrate the onset of a DSW. This behaviour is intriguing as, unlike classical statistical mechanical systems, e.g. magnetic and fluid models \cite{baxter}, the order parameter $u(x)$ develops oscillations in the form of a dispersive shock in conjunction with the existence of additional stationary points for the free energy such as unstable and metastable states. \textcolor{black}{The emergence of such oscillations is therefore explained as a result of higher order corrections in the string equation \eqref{stringformal} of which the cubic equation \eqref{interpu} is the leading order approximation. Based on the results in~\cite{dubrovin3a,claeys,dubrovin3,dgkm}, the mechanism of formation of such oscillations in the vicinity of the of the critical points of the solution to the equation~\eqref{interpu} is universal and it is given by a particular real analytic solution of the second member of the Painlev\'e I hierarchy, known as  Painlev\'e I2 equation.}

\section{Analysis of scenarios} %
  \begin{figure}[t]
  \begin{center}
  \begin{subfigure}[b]{0.35\textwidth}
    \includegraphics[width=\textwidth, frame]{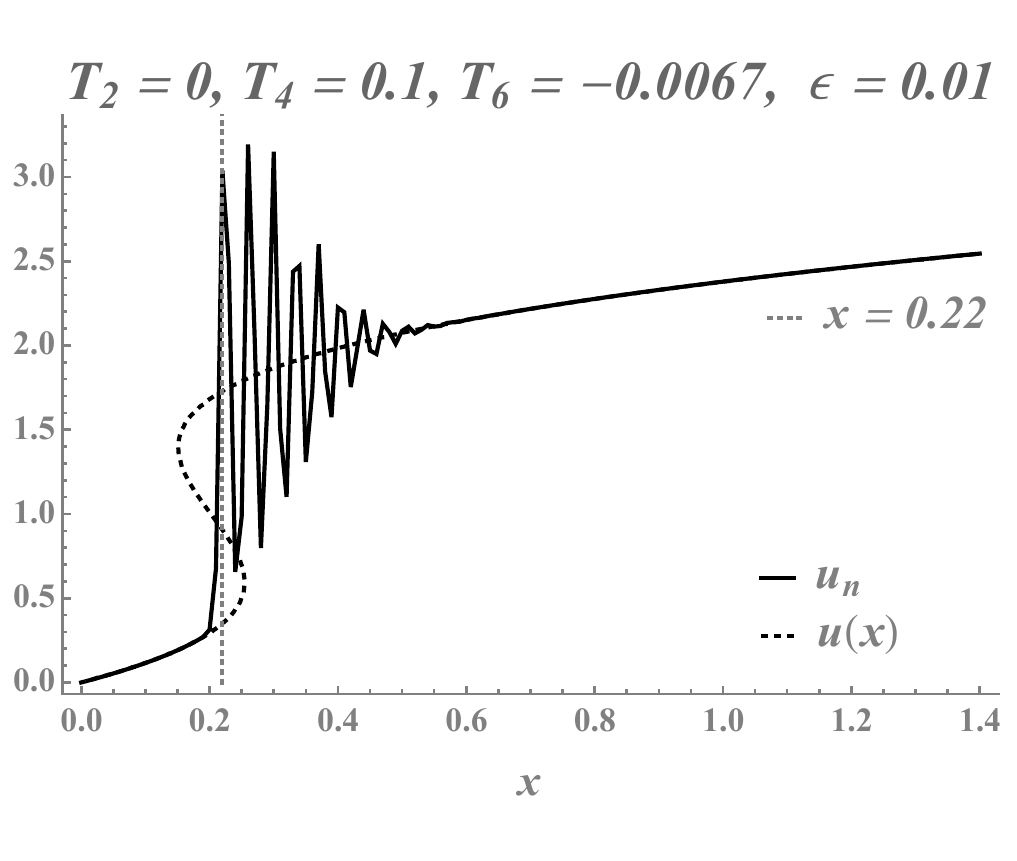}\caption{}
   \label{fig:mixC}
  \end{subfigure}
 \quad
  \begin{subfigure}[b]{0.35\textwidth}
    \includegraphics[width=\textwidth, frame]{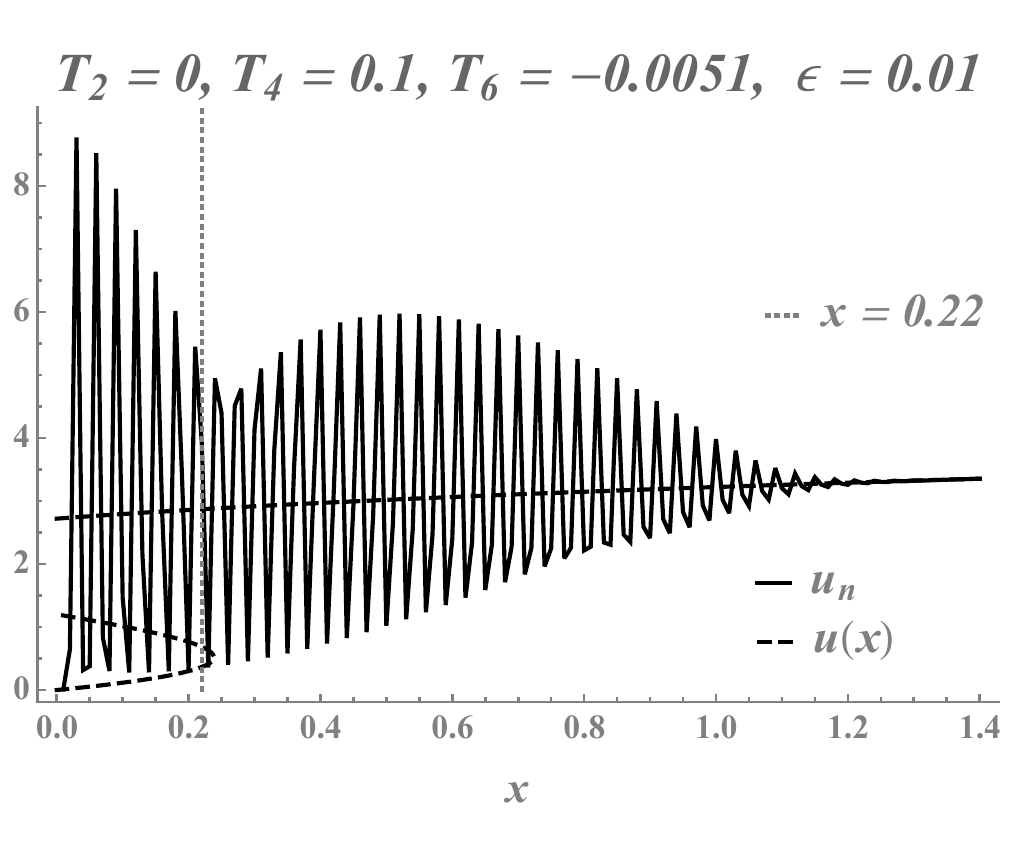}\caption{}
    \label{fig:mixD}
  \end{subfigure}
  \end{center}
\caption{\small In all figures $T_2 = 0$ and $T_4 = 0.1$. \textcolor{black}{ Fig.s \ref{fig:mixC} and \ref{fig:mixD} show a comparison of the order parameter evaluated  by using Eq. \eqref{eqrecurrenceU} and Eq. \eqref{interpu} at $T_6 =-0.0067$ and $T_6 =-0.0051$ respectively. The choice of values of the coupling constants  coincide with the ones in Fig. \ref{fig:mixB} and the dotted vertical line marks the value $x=0.22$ where the free energy density shows two local minima.}}
  \label{fig:mix}
\end{figure}
\textcolor{black}{As discussed above, a dispersive shock in the order parameter develops for values of the couplings such that the density of free energy admits multiple stationary points. Hence, the signature of the discriminant of the cubic in Eq. \eqref{interpu} determines a necessary condition for the occurrence of a dispersive shock and therefore a phase transition. We now focus on the different scenarios in regimes where the discriminant $\Delta$ is positive and the Eq. \eqref{interpu} admits three real and distinct solutions. In particular, different cases need to be considered depending on whether the coefficients of the cubic equation \eqref{interpu} are negative or positive.}
Necessarily, in order to ensure convergence of the integral \eqref{partition}, it is $T_{6} < 0$. Hence, we have four distinct cases, depending on the signs of the coefficient $1- 2T_{2}$ and $-12T_{4}$ in Eq.~\eqref{interpu}.\\

\begin{figure}[t]
\begin{center}
\begin{subfigure}[b]{0.35\textwidth}
    \includegraphics[width=\textwidth, frame]{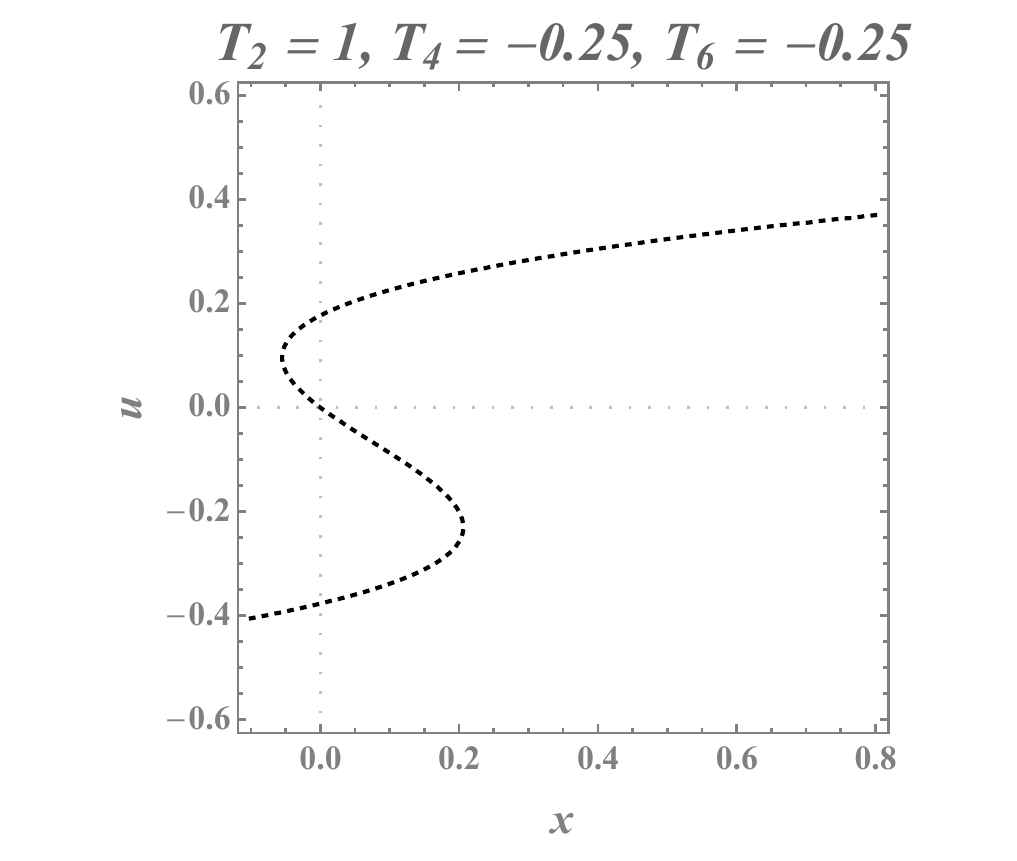}  \caption{}
    \label{fig:mix2A}
  \end{subfigure}
 \quad
  \begin{subfigure}[b]{0.35\textwidth}
    \includegraphics[width=\textwidth, frame]{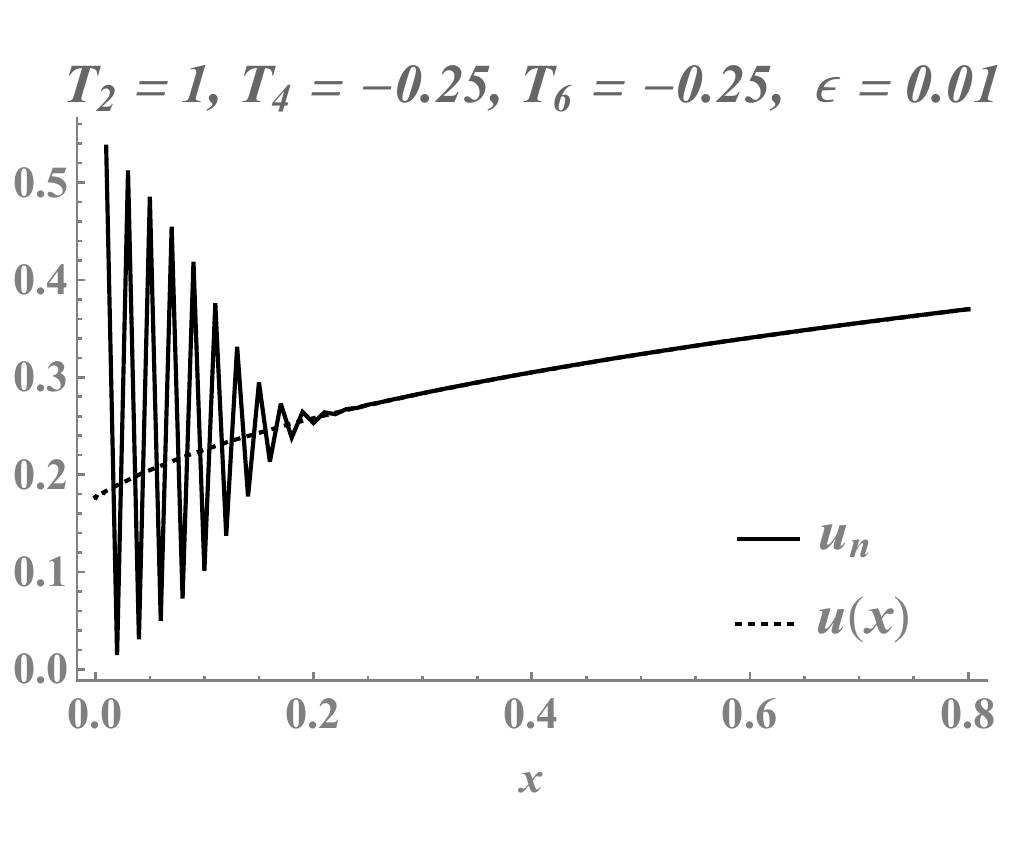}\caption{}
    \label{fig:mix2B}
  \end{subfigure}\\
  \end{center}
  \caption{\small \textcolor{black}{In all figures $T_2 =1$, \textcolor{black}{$T_4 = -0.25$} and $T_6 = -0.25$. Fig.~\ref{fig:mix2A} shows the multivalued solution of Eq.~\eqref{interpu} to be compared with the order parameter evaluated via Eq.~\eqref{eqrecurrenceU} and Eq. \eqref{interpu} and shown in  Fig. \ref{fig:mix2B}. The catastrophe occurs at $x<0$ and generates a dispersive shock propagating across the origin and for $x>0$.}}
  \label{fig:mix2}
  \end{figure}

\noindent \textcolor{black}{{\bf Scenario 1: $T_2 < 1/2$ and $T_4 > 0$.}} \\ This choice corresponds to the case analysed in~\cite{jurkiewicz,jurkiewicz2}, hence it allows for a direct comparison. \textcolor{black}{As by definition $u(x)\geq 0$,  only non negative branches of $u(x)$ correspond to admissible states of the system. In fact, in both Figs. \ref{fig:mixC} and \ref{fig:mixD}, the three branches of the cubic which correspond to stationary points of the free energy are positive.}\\

 \begin{figure}[t]
  \begin{center}
  \begin{subfigure}[b]{0.35\textwidth}
    \includegraphics[width=\textwidth, frame]{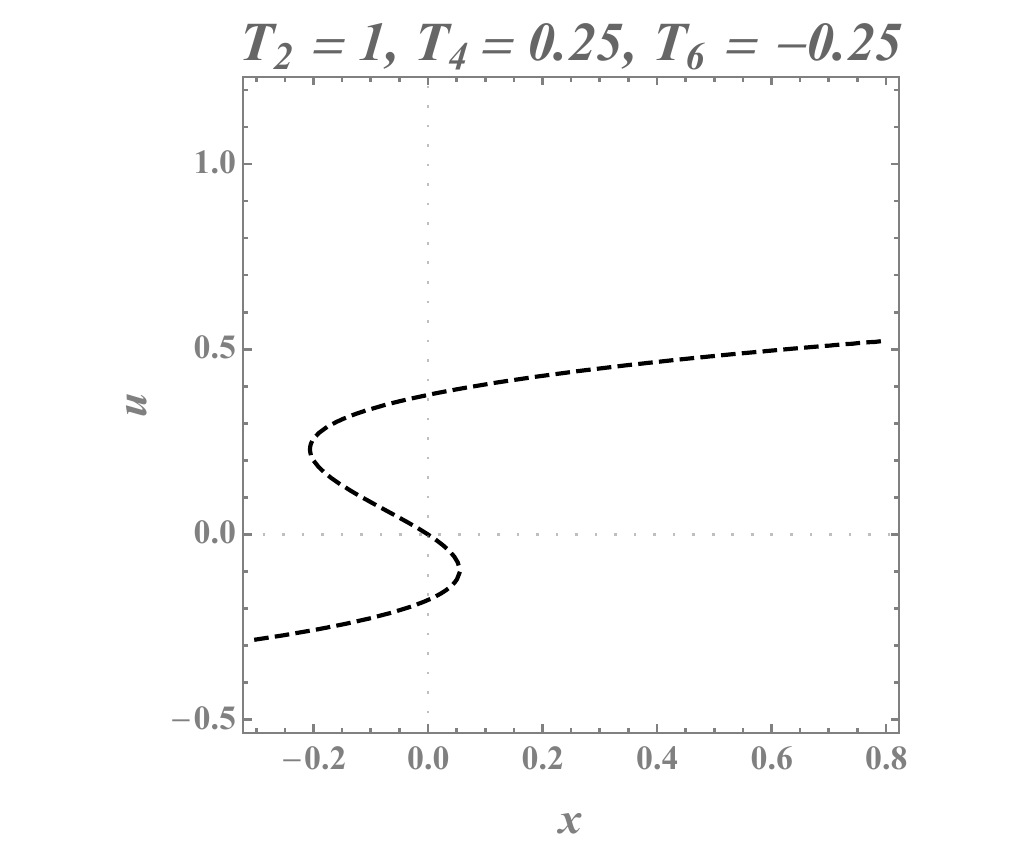}  \caption{}
    \label{fig:mix2C}
  \end{subfigure}
 \quad
  \begin{subfigure}[b]{0.35\textwidth}
    \includegraphics[width=\textwidth, frame]{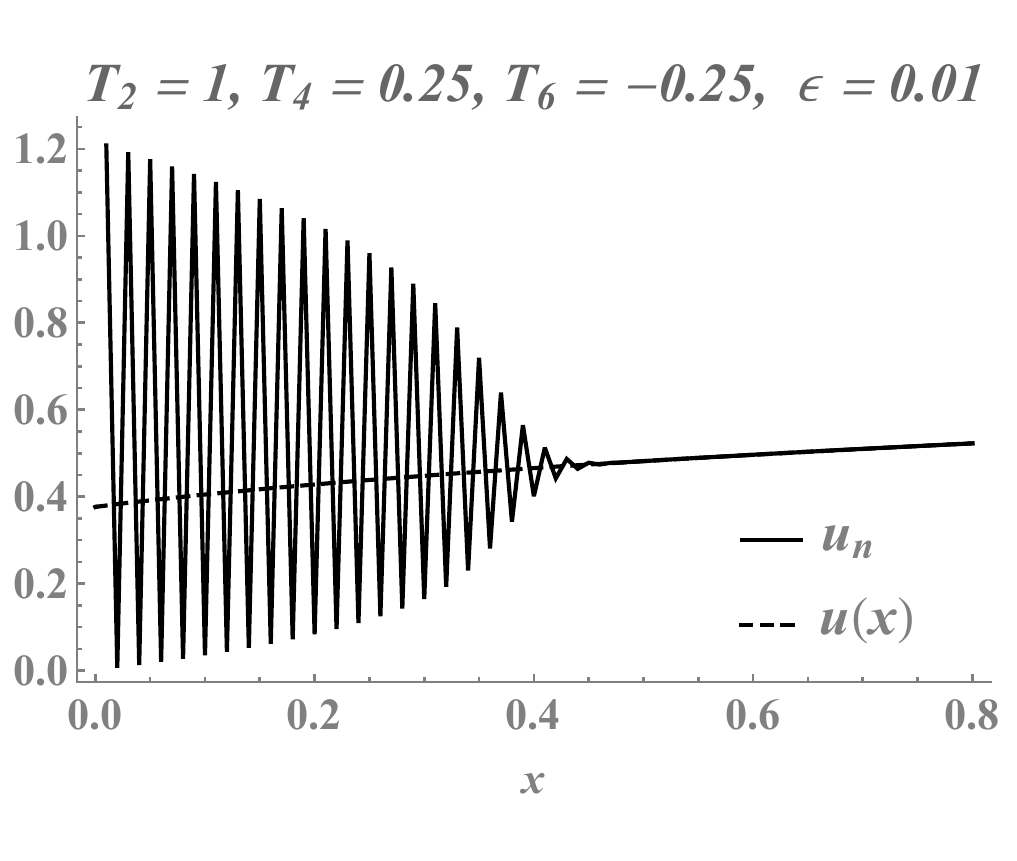}\caption{}
    \label{fig:mix2D}
  \end{subfigure}
\end{center}
 \caption{\small \textcolor{black}{In all figures $T_2 =1$, \textcolor{black}{$T_4 = 0.25$} and $T_6 = -0.25$. Fig.~\ref{fig:mix2A} shows a multivalued solution of Eq.~\eqref{interpu} to be compared with the order parameter evaluated by using Eq.~\eqref{eqrecurrenceU} and Eq. \eqref{interpu}  and shown in Fig. \ref{fig:mix2B}. The dispersive shock generates from a gradient catastrophe at $x>0$.}}
  \label{fig:mix2bis}
\end{figure}

\noindent  \textcolor{black}{{\bf Scenario 2: $T_2 > 1/2$ and $T_4<0$.} }  \\ 
\textcolor{black}{In this case, as illustrated for example in Fig. \ref{fig:mix2}}, the solution to Eq.~\eqref{interpu}, \textcolor{black}{shown in  Fig. \ref{fig:mix2A}}, it is three-valued but one root, associated to a local minimum, is negative and therefore does not correspond to a state of the system. However, two concurrent states, although of different nature, one stable and one unstable, underlie a dispersive shock, shown in Fig. \ref{fig:mix2B}. Notice that for $x>0$ the solution to Eq.~\eqref{interpu} has one non negative branch only. 
Nonetheless, $u(x)$  develops a dispersive shock profile at positive $x$, although this is originated by a catastrophe located at $x<0$.  The solution to Eq.~\eqref{interpu} is multivalued with two non negative branches for a small interval of negative values of $x$. \\

\noindent  \textcolor{black}{{\bf Scenario 3: $T_2>1/2$ and  $T_4>0$.}} \textcolor{black}{Similarly to Case 2, $u(x)$ has only one positive solution  for $x>0$ (see Fig. \ref{fig:mix2C}) and as shown in Fig. \ref{fig:mix2D}, a dispersive shock arises in correspondence of two non negative roots for negative $x$.}

\begin{figure}[t]
\begin{center}
\begin{subfigure}[b]{0.30\textwidth}
    \includegraphics[width=\textwidth, frame]{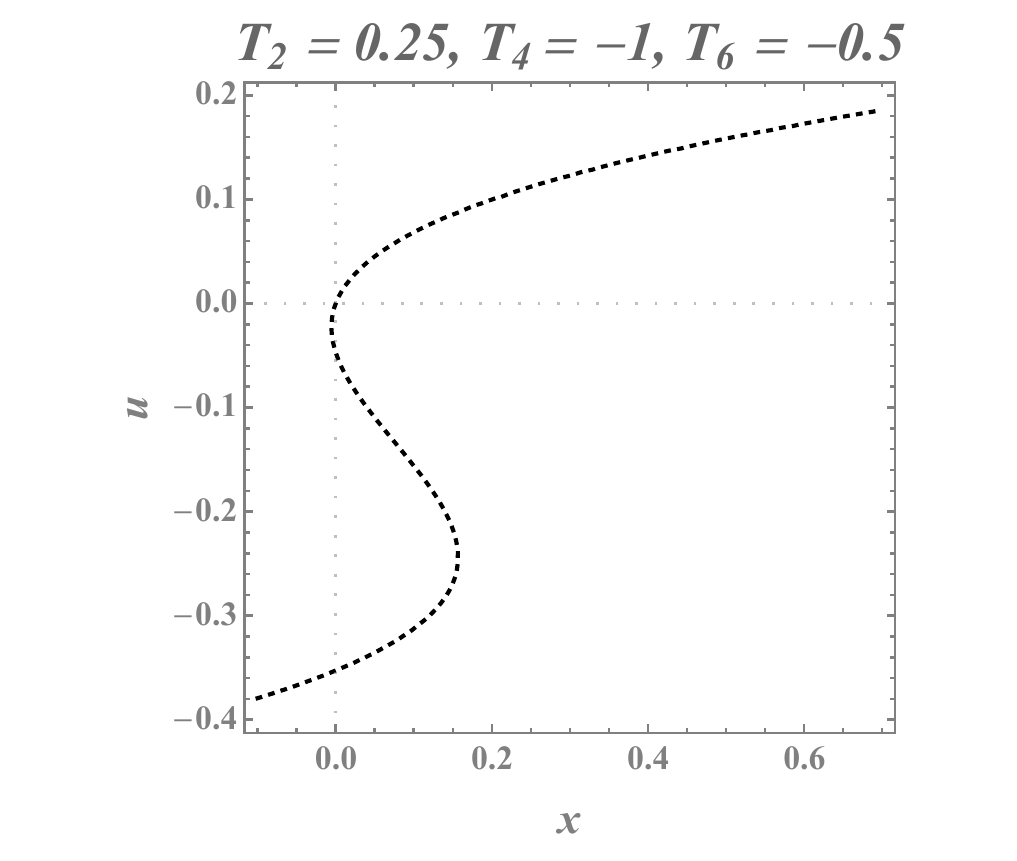}  \caption{}
    \label{fig:scaleA}
  \end{subfigure}
 \quad
  \begin{subfigure}[b]{0.30\textwidth}
    \includegraphics[width=\textwidth, frame]{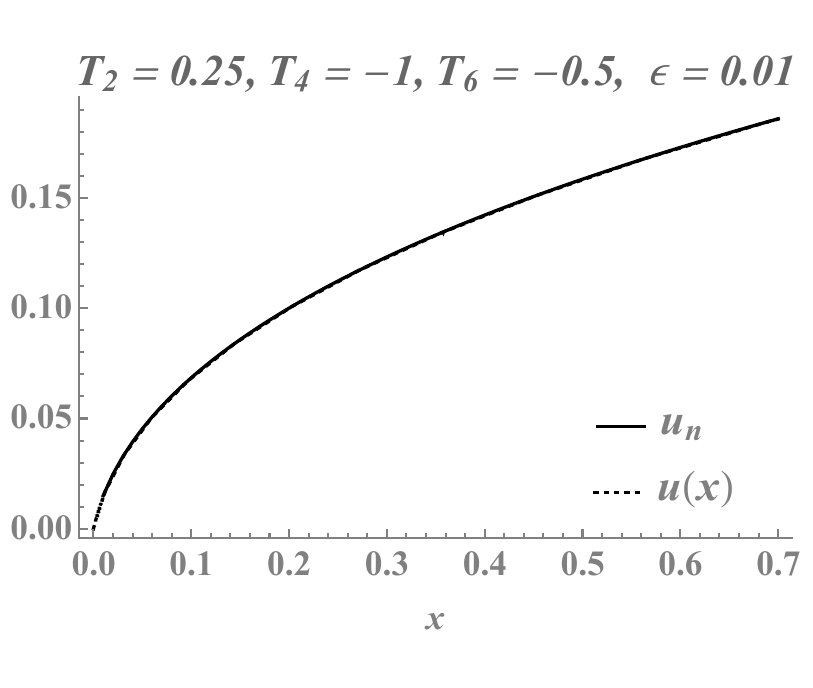}\caption{}
    \label{fig:scaleB}
  \end{subfigure}
\end{center}
\caption{ {\small \textcolor{black}{In all figures $T_2 = 0.25$, $T_4 = -1$, $T_6 = -0.5$. Fig. \ref{fig:scaleA} shows the solution of Eq.~\eqref{interpu}. Fig. \ref{fig:scaleB} shows a comparison of the solution of Eq.~\eqref{eqrecurrenceU} with the non negative branch of  the solution to Eq.~\eqref{interpu}. In this case, the presence of only one local minimum of the free energy density, where the order parameter is positive, suppresses the formation of a dispersive shock.}
}}
\label{fig:onepos}
\end{figure}

\noindent \textcolor{black}{{\bf Scenario 4: $T_2 <1/2$ and $T_4<0$.} The last scenario is given by the case, illustrated in Fig. \ref{fig:scaleA}, where the cubic solution is multivalued with one positive branch for all values of $x$ and therefore only one solution corresponds to a state that is accessible by the system. Interestingly, as shown in Fig. \ref{fig:scaleB}}, the solution of the recurrence equation~\eqref{eqrecurrenceU} overlaps with the cubic solution and no oscillations occur. \textcolor{black}{This suggests} that the dispersive regularisation in the form of a dispersive shock of the order parameter is related to the existence of {\it accessible} \mbox{(meta-)stable/unstable} states. \textcolor{black}{This phenomenon is rather unexpected as,} from the point of view of the governing dispersive PDE, one would expect that for a generic initial condition multivaluedness would be replaced by a dispersive shock originated by the gradient catastrophe. 
\section{Conclusions}
We have analysed the Hermitian matrix model with even degree of nonlinearity for which the order parameter of the model is obtained as a particular solution of the Volterra hierarchy. \textcolor{black}{The critical properties of the model are obtained in terms of a solution to the continuum limit of Volterra hierarchy.} We have shown that the chaotic behaviour  observed in previous literature can be described as a dispersive shock propagating in the space of thermodynamic parameters. Also, the profile of the order parameter, specifically the form of the envelope, appears to be highly sensitive to the choice of the parameters $T_{2k}$. For instance, Fig.s~\ref{fig:mixC},  \ref{fig:mixD}, \ref{fig:mix2B} and~\ref{fig:mix2D} show the onset of dispersive shocks whose envelope displays very distinctive features.

A further detailed study of this intriguing behaviour will involve the construction of the asymptotic genus expansion  of the solution~\eqref{eqrecurrenceU}  and the Whitham modulation theory for solutions of related integrable hierarchy. We finally \textcolor{black}{anticipate that the rich phenomenology described in this paper reflects the fact that the dispersive shock given by the solution~\eqref{eqrecurrenceU} is an intrinsic multidimensional object arising from the simultaneous solution of equations of the hierarchy~\eqref{Todareduced} in the continuum limit.  In fact, the solution to the Volterra hierarchy~\eqref{Todareduced} satisfies  the modified Kadomtsev-Petviashvili hierarchy similarly to how the solution of the Toda hierarchy satisfies the Kadomtsev-Petviashvili hierarchy. Hence, the description of the dispersive shock solution entails the development of the Whihtam modulation theory of a $2+1$-dimensional integrable dispersive equation. This further development is in progress and will be presented elsewhere~\cite{BBMP}.}

\textbf{Acknowledgements.} This work is dedicated to the memory of Boris Dubrovin (1950-2019) whose magnificent scientific foresight and generosity have inspired the original ideas of this research. Authors are grateful to The Leverhulme Trust RPG-2017-228 {\it Integrable Dressed Networks} for supporting this research project.

\end{document}